\documentclass[aps,prl,showpacs,showkeys,amsmath,amssymb,
twocolumn,
floatfix,
superscriptaddress
]{revtex4-1}
\usepackage{graphicx}
\usepackage{times}

\newcommand{\nuebar}{$\overline{\nu}_{e}$}

\begin{document}

\title{Observation of electron-antineutrino disappearance at Daya Bay}

%\author[]{Daya~Bay~Collaboration}

\newcommand{\IHEP}{\affiliation{Institute~of~High~Energy~Physics, Beijing}}
\newcommand{\USTC}{\affiliation{University~of~Science~and~Technology~of~China, Hefei}}
\newcommand{\UW}{\affiliation{University~of~Wisconsin, Madison, WI}}
\newcommand{\BNL}{\affiliation{Brookhaven~National~Laboratory, Upton, NY}}
\newcommand{\NUU}{\affiliation{National~United~University, Miao-Li}}
\newcommand{\CIT}{\affiliation{California~Institute~of~Technology, Pasadena, CA}}
\newcommand{\NJU}{\affiliation{Nanjing~University, Nanjing}}
\newcommand{\THU}{\affiliation{Department of Engineering Physics, Tsinghua~University, Beijing}}
\newcommand{\CUHK}{\affiliation{Chinese~University~of~Hong~Kong, Hong~Kong}}
\newcommand{\SZU}{\affiliation{Shenzhen~Univeristy, Shen~Zhen}}
\newcommand{\Siena}{\affiliation{Siena~College, Loudonville, NY}}
\newcommand{\IIT}{\affiliation{Department of Physics, Illinois~Institute~of~Technology, Chicago, IL}}
\newcommand{\LBNL}{\affiliation{Lawrence~Berkeley~National~Laboratory, Berkeley, CA}}
\newcommand{\UIUC}{\affiliation{Department of Physics, University~of~Illinois~at~Urbana-Champaign, Urbana, IL}}
\newcommand{\CDUT}{\affiliation{Chengdu~University~of~Technology, Chengdu}}
\newcommand{\JINR}{\affiliation{Joint~Institute~for~Nuclear~Research, Dubna, Moscow~Region}}
\newcommand{\SJTU}{\affiliation{Shanghai~Jiao~Tong~University, Shanghai}}
\newcommand{\BNU}{\affiliation{Beijing~Normal~University, Beijing}}
\newcommand{\PU}{\affiliation{Joseph Henry Laboratories,~Princeton~University, Princeton, NJ}}
\newcommand{\NTU}{\affiliation{Department~of~Physics, National~Taiwan~University, Taipei}}
\newcommand{\VT}{\affiliation{Center~for~Neutrino~Physics, Virginia~Tech, Blacksburg, VA}}
\newcommand{\NCTU}{\affiliation{Institute~of~Physics, National~Chiao-Tung~University, Hsinchu}}
\newcommand{\CIAE}{\affiliation{China~Institute~of~Atomic~Energy, Beijing}}
\newcommand{\UCLA}{\affiliation{University~of~California,~Los~Angeles, CA}}
\newcommand{\UH}{\affiliation{Department~of~Physics, University~of~Houston, Houston, TX}}
\newcommand{\SDU}{\affiliation{Shandong~University, Jinan}}
\newcommand{\NU}{\affiliation{School~of~Physics, Nankai~University, Tianjin}}
\newcommand{\UC}{\affiliation{Department of Physics, University~of~Cincinnati, Cincinnati, OH}}
\newcommand{\DGUT}{\affiliation{Dongguan~Institute~of~Technology, Dongguan}}
\newcommand{\UCB}{\affiliation{Department of Physics, University~of~California, Berkeley, CA}}
\newcommand{\UHK}{\affiliation{Department~of~Physics, The University of Hong Kong, Pokfulam, Hong Kong}}
\newcommand{\CU}{\affiliation{Charles University, Faculty of Mathematics and Physics, Prague}}
\newcommand{\SYSU}{\affiliation{Sun~Yat-Sen~(Zhongshan)~University, Guangzhou}}
\newcommand{\WM}{\affiliation{College~of~William~and~Mary, Williamsburg, VA}}
\newcommand{\RPI}{\affiliation{Department of Physics, Applied Physics, and Astronomy, Rensselaer~Polytechnic~Institute, Troy, NY}}
\newcommand{\CGNPHC}{\affiliation{China~Guangdong~Nuclear~Power~Group, Shenzhen}}
\newcommand{\ISU}{\affiliation{Iowa~State~University, Ames, IA}}
\newcommand{\NCEPU}{\affiliation{North China Electric Power University, Beijing}}

\author{F.~P.~An}\IHEP
\author{J.~Z.~Bai}\IHEP
\author{A.~B.~Balantekin}\UW
\author{H.~R.~Band}\UW
\author{D.~Beavis}\BNL
\author{W.~Beriguete}\BNL
\author{M.~Bishai}\BNL
\author{S.~Blyth}\NUU
\author{K.~Boddy}\CIT
\author{R.~L.~Brown}\BNL
% ...

\collaboration{The Daya Bay Collaboration}\noaffiliation
\date{\today}

\begin{abstract}
\noindent
The Daya Bay Reactor Neutrino Experiment has measured a non-zero value for
the neutrino mixing angle $\theta_{13}$ with a significance of 5.2 standard deviations.
Antineutrinos from six 2.9 GW$_{\rm th}$ reactors were detected in six antineutrino detectors deployed in two near (flux-weighted baseline 470 m and 576 m) and one far (1648 m) underground experimental halls.
With a 43,000 ton-GW$_{\rm th}$-day livetime exposure in 55 days, 10416 (80376) electron antineutrino candidates were detected at the far hall (near halls). The ratio of the observed to expected number of antineutrinos at the far hall is $R=0.940\pm 0.011({\rm stat}) \pm 0.004({\rm syst})$.
A rate-only analysis finds $\sin^22\theta_{13}=0.092\pm 0.016({\rm stat})\pm0.005({\rm syst})$ in a three-neutrino framework.
\end{abstract}

\maketitle

\par
It is well established that the flavor of a neutrino oscillates with time. Neutrino oscillations can be described by the three mixing angles ($\theta_{12}$, $\theta_{23}$, and $\theta_{13}$) and a phase of the
Pontecorvo-Maki-Nakagawa-Sakata matrix, and two mass-squared
differences ($\Delta m^2_{32}$ and $\Delta m^2_{21}$)~\cite{pontecorvo,mns}.
Of these mixing angles, $ \theta_{13}$ is the least
known. The CHOOZ experiment obtained a 90\%-confidence-level upper limit of 0.17 for
sin$^22\theta_{13}$~\cite{chooz}.
Recently, results from T2K~\cite{t2k}, MINOS~\cite{minosth13} and Double Chooz~\cite{dchooz}
have indicated that $\theta_{13}$ could be non-zero.
In this paper, we present the observation of a non-zero value for $\theta_{13}$.

\par
For reactor-based experiments, an unambiguous determination of $\theta_{13}$
can be extracted via the survival probability of the electron antineutrino \nuebar\ at short distances from the reactors,
\begin{equation}\label{eqn:psurv}
P_{\rm sur} \approx 1 - \sin^2 2\theta_{13} \sin^2 (1.267 \Delta m^2_{31}  L/E) \,,
\end{equation}
where $\Delta m^2_{31}=\Delta m^2_{32}\pm\Delta m^2_{21}$,
$E$ is the \nuebar\ energy in MeV and $L$ is the distance in meters between
the \nuebar\ source and the detector (baseline).

\par
\begin{figure}[htb]
\includegraphics[width=\columnwidth]{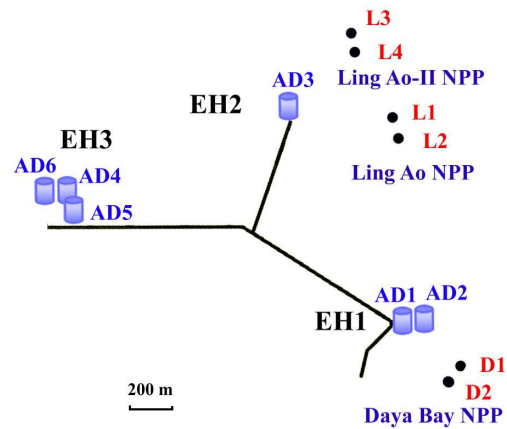}
\caption{Layout of the Daya Bay experiment.
The dots represent reactors, labeled as D1, D2, L1, L2, L3 and L4.
Six ADs, AD1--AD6, are installed in three EHs. \label{fig:layout}}
\end{figure}

The near-far arrangement of antineutrino detectors (ADs), as illustrated in Fig.~\ref{fig:layout}, allows for a relative
measurement by comparing the observed \nuebar\ rates at various baselines.
With functionally identical ADs, the relative rate is independent of correlated uncertainties and uncorrelated reactor uncertainties are minimized.

\par
A detailed description of the Daya Bay experiment can be found in~\cite{ad12,dyb}.
Here, only the apparatus relevant to this analysis will be highlighted.
The six pressurized water reactors are grouped into three pairs with
each pair referred to as a nuclear power plant (NPP)\@.
The maximum thermal power of each reactor is 2.9~GW$_{\rm th}$\@.
Three underground experimental halls (EHs) are
connected with horizontal tunnels. Two ADs are located in EH1 and one
in EH2 (the near halls). Three ADs are positioned near the oscillation maximum in the far hall, EH3. The vertical overburden in equivalent meters of water (m.w.e.), the simulated muon rate and average muon energy, and average distance to the reactor pairs are listed in
Table~\ref{tab:baseline}.

\begin{table}[htb]
\begin{center}
\begin{tabular}[c]{cccrrrr} \hline\hline
& Overburden & \mbox{ }$R_\mu$ &\mbox{ } $E_\mu$ & \mbox{  }D1,2 &\mbox{  } L1,2 & \mbox{   }L3,4 \\\hline
EH1 & 250 & 1.27 & 57 & 364 & 857 & 1307\\
EH2 & 265 & 0.95 & 58 & 1348 & 480 & 528\\
EH3 & 860 & 0.056 & 137 &\mbox{  }  1912 &\mbox{  }  1540 &\mbox{  }  1548\\\hline
\end{tabular}
\caption{Vertical overburden (m.w.e.), muon rate $R_\mu$ (Hz/m$^2$),
and average muon energy $E_\mu$ (GeV) of the three EHs,
and the distances (m) to the reactor pairs.\label{tab:baseline}}
\end{center}
\end{table}

\par
\begin{figure}[htb]
\includegraphics[width=\columnwidth]{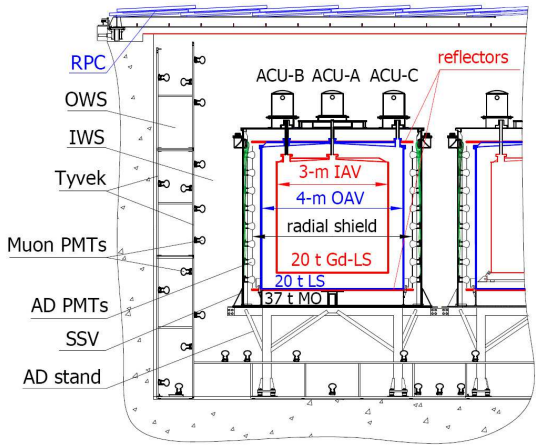}
\caption{Schematic diagram of the Daya Bay detectors. \label{fig:det}}
\end{figure}

As shown in Fig.~\ref{fig:det}, the ADs in each EH are shielded with $>$2.5~m of high-purity water against ambient radiation in all directions.  Each water pool is segmented into inner and outer water shields (IWS and OWS) and instrumented with photomultiplier tubes (PMTs) to function as Cherenkov-radiation detectors whose data were used by offline software to remove spallation neutrons and other cosmogenic backgrounds.  The detection efficiency for long-track muons is $>$99.7\%~\cite{ad12}.

\par
The \nuebar\ is detected via the inverse $\beta$-decay (IBD) reaction,
$\overline{\nu}_e + p \to e^+ + n$, in a Gadolinium-doped liquid scintillator
(\mbox{Gd-LS})~\cite{ding,yeh}\@. The coincidence of the prompt scintillation from the $e^+$ and the
delayed neutron capture on Gd provides a distinctive \nuebar\ signature.

\par
Each AD consists of a cylindrical, \mbox{5-m} diameter stainless steel vessel (SSV) that houses two nested, UV-transparent acrylic cylindrical vessels.  A \mbox{3.1-m} diameter inner acrylic vessel (IAV) holds 20 t of \mbox{Gd-LS} (target). It is surrounded by a region with 20 t of liquid scintillator (LS) inside a \mbox{4-m} diameter outer acrylic vessel (OAV)\@.  Between the SSV and OAV, 37 t of mineral oil (MO) shields the LS and \mbox{Gd-LS} from radioactivity. IBD interactions are detected by 192 Hamamatsu R5912 PMTs. A black radial shield and specular reflectors are installed on the vertical detector walls and above and below the LS volume, respectively. \mbox{Gd-LS} and LS are prepared and filled into ADs systematically to ensure all ADs are functionally identical~\cite{ad12}.
Three automated calibration units (ACUs) mounted on the SSV lid allow for remote
deployment of an LED, a $^{68}$Ge source, and a combined source of
\mbox{$^{241}$Am-$^{13}$C} and $^{60}$Co into the \mbox{Gd-LS} and LS liquid volumes along
three vertical axes.

\par
The results are based on data taken from 24 December 2011 to 17 February
2012.  A blind analysis strategy was adopted, with the baselines, the thermal
power histories of the cores, and the target masses of the ADs hidden until the analyses were frozen.
Triggers were formed from the number of PMTs with signals above a $\sim$0.25 photoelectron (pe) threshold
(NHIT) or the charge-sum of the over-threshold PMTs (ESUM)\@.
The AD triggers were NHIT $>$ 45 or ESUM $\gtrsim$ 65 pe.
The trigger rate per AD was $<\!280$ Hz with a negligible trigger inefficiency for IBD candidates.
The data consist of charge and timing information for each PMT, and were accumulated independently for each detector.
To remove systematic effects due to reactor flux fluctuations, only data sets with all detectors in operation were used.

\par
The energy of each trigger in an AD was reconstructed based on the total pe collected by the PMTs.  The energy calibration constant, $\sim$163 pe/MeV for all ADs and stable throughout the
data collection period, was determined by setting the energy peak of the $^{60}$Co source deployed at each AD
center to 2.506 MeV\@.  Vertex reconstruction was based on
center-of-charge (COC), defined as the charge-weighted-mean of the
coordinates of all PMTs. The mapping from COC to vertex was done by
analytic corrections determined using data collected with $^{60}$Co sources deployed at various points within the AD\@.
A vertex-dependent correction to energy ($<$10\%) and a constant factor (0.988) were applied equally to all ADs to correct for geometrical effects and energy nonlinearity between the  $^{60}$Co and the neutron capture on Gd ($n$Gd), determined by the  $^{60}$Co and \mbox{Am-C} sources at the detector center.
An independent energy calibration that utilized the peak of the $n$Gd from spallation neutron to set the energy scale and templates derived from Monte Carlo simulations (MC) for vertex
reconstruction, gave consistent performance~\cite{ad12}. The energy resolution was
(7.5/$\sqrt{E({\rm MeV})}+0.9)\%$ for all 6 ADs.

\par
IWS and OWS triggers with NHIT $>\!12$ were
classified as `WS muon candidates' or $\mu_{\rm WS}$.  Events in an AD within $\pm2$~$\mu$s of a $\mu_{\rm WS}$ with energy $>$20 MeV and $>$2.5 GeV were classified as muons ($\mu_{\rm AD}$) and showering muons ($\mu_{\rm sh}$), respectively, for vetoing purposes.
An instrumental background due to spontaneous light
emission from a PMT, denoted as a flasher, was rejected efficiently~\cite{ad12}.

\par
IBD events were selected with the following criteria:
 $0.7 \!<\! E_p \!<\! 12.0$ MeV,
 $6.0 \!<\! E_d \!<\! 12.0$ MeV,
 $1   \!<\! \Delta t \!<\! 200$ $\mu$s,
 the prompt-delayed pair was vetoed by preceding muons if $t_d - t_{\mu_{\rm WS}} \!<\! 600\ \mu$s,
$t_d - t_{\mu_{\rm AD}} \!<\! 1000\ \mu$s or $t_d - t_{\mu_{\rm sh}} \!<\! 1$ s,
and
 a multiplicity cut that requires no additional $>$0.7~MeV trigger in the time range
$(t_p-200\mu{\rm s}, t_d+200\mu{\rm s})$,
where $E_p$ ($E_d$) is the prompt (delayed)
energy and $\Delta t = t_d - t_p$ is the time
difference between the prompt and delayed signals.
Statistically consistent performance was achieved by an independent analysis that used
different energy reconstruction, muon veto, and multiplicity cuts.

\par
The inefficiency of the muon veto for selecting IBD events $(1-\epsilon_{\mu})$ was calculated by integrating the vetoed time of each muon with temporal overlaps taken into account. Inefficiency due to the multiplicity selection $(1-\epsilon_{m})$ was calculated by considering the probability that a random signal occurred
near an IBD in time. The average values of $\epsilon_\mu\cdot\epsilon_m$ are given for each AD in Table~\ref{tab:ibd}.

\begin{table*}[!htb]
\begin{tabular}{|c|cc|c|ccc|}
\hline
                  & AD1  & AD2  & AD3 & AD4 & AD5 & AD6 \\
\hline
IBD candidates &  28935  &  28975  &  22466 &  3528  &  3436  &  3452 \\
\hline
No-oscillation prediction for IBD & 28647 & 29096 & 22335 & 3566.5 & 3573.0 & 3535.9 \\
\hline
DAQ live time (days) & \multicolumn{2}{c|}{49.5530}     &  49.4971  &    &   48.9473  &    \\
\hline
Muon veto time (days) &  8.7418  &  8.9109  &  7.0389  &  0.8785  &  0.8800  & 0.8952  \\
\hline
$\epsilon_{\mu}\cdot\epsilon_{m}$ &  0.8019 &  0.7989  &  0.8363  & 0.9547  &  0.9543  &  0.9538 \\
\hline
Accidentals (per day) &  9.82$\pm$0.06  &  9.88$\pm$0.06  & 7.67$\pm$0.05   &  3.29 $\pm$0.03  &  3.33 $\pm$ 0.03  &  3.12 $\pm$0.03 \\
\hline
Fast-neutron (per day) &  0.84$\pm$0.28  &  0.84$\pm$0.28   &  0.74$\pm$0.44   &  0.04$\pm$0.04  &
 0.04$\pm$0.04 & 0.04$\pm$0.04 \\
\hline
$^9$Li/$^8$He (per AD per day) &  \multicolumn{2}{c|}{3.1$\pm$1.6}  & 1.8$\pm$1.1 & \multicolumn{3}{c|}{0.16$\pm$0.11}   \\
\hline
Am-C correlated (per AD per day) &  \multicolumn{6}{c|}{0.2$\pm$0.2}   \\
\hline
 $^{13}$C($\alpha$, n)$^{16}$O background (per day)&  0.04$\pm$0.02  &  0.04$\pm$0.02  & 0.035$\pm$0.02 & 0.03$\pm$0.02  & 0.03$\pm$0.02 & 0.03$\pm$0.02   \\
\hline
IBD rate (per day) &  714.17$\pm$4.58  & 717.86$\pm$ 4.60 & 532.29$\pm$3.82  & 71.78 $\pm$ 1.29  & 69.80$\pm$1.28  & 70.39$\pm$1.28 \\
\hline
\end{tabular}
\caption{Signal and background summary. The background and IBD rates were corrected for the $\epsilon_{\mu}\cdot\epsilon_{m}$ efficiency. The no-oscillation predictions based on reactor flux analyses and detector simulation have been corrected with the best-fit normalization parameter in determining $\sin^22\theta_{13}$.\label{tab:ibd}  }
\end{table*}

\par
We considered the following kinds of background: accidental correlation of two unrelated signals,
$\beta$-n decay of $^9$Li/$^8$He produced by muons in the ADs, fast-neutron backgrounds produced by muons outside the ADs, $^{13}$C($\alpha$,n)$^{16}$O interactions, and
correlated events due to the retracted \mbox{Am-C} neutron source in the ACUs.
The estimated background rates per AD are summarized in Table~\ref{tab:ibd}.

\par
The accidental background was determined by measuring the
rate of both prompt- and delayed-like signals, and then estimating
the probability that two signals randomly satisfied the $\Delta$t required for IBD
selection. Additional estimates using prompt and delayed candidates separated by more than 1 ms or 2 meters provided consistent results. The uncertainty in the measured accidental rate was dominated by
the statistical uncertainty in the rate of delayed candidates.

\par
The rate of correlated background from the $\beta$-$n$ cascade of
$^9$Li/$^8$He decays was evaluated from the distribution of the time since the last
muon using the known decay times for these isotopes~\cite{wenljnim}.
The $^9$Li/$^8$He background rate as a function of the muon energy deposited in the
AD was estimated by preparing samples with and without detected neutrons 10 $\mu$s to 200 $\mu$s after the muon. A 50\% systematic uncertainty was assigned to account for the extrapolation to zero deposited muon energy.

\par
An energetic neutron entering an AD can form a fast-neutron background by recoiling off
a proton before being captured on Gd.  By relaxing the $E_p\!<\!12$ MeV criterion in the IBD selection, a flat distribution in $E_p$ was observed up to 100 MeV\@. Extrapolation into the IBD energy region gave an estimate for the residual fast-neutron background.
A similar flat $E_p$ distribution
was found in the muon-tagged fast-neutron sample produced by inverting the muon veto cut. Consistent results were obtained by scaling the muon-tagged fast-neutron rate with muon inefficiency, and by MC\@.

\par
The $^{13}$C($\alpha$,n)$^{16}$O background was determined using MC after estimating the amount of $^{238}$U, $^{232}$Th, $^{227}$Ac, and $^{210}$Po in the \mbox{Gd-LS} from their cascade decays, or by fitting their $\alpha$-particle energy peaks in the data.

\par
A neutron emitted from the 0.5-Hz \mbox{Am-C} neutron source in an ACU could
generate a gamma-ray via inelastic scattering in the SSV before subsequently being captured on Fe/Cr/Mn/Ni.
An IBD was mimicked if both gamma-rays from the scattering
and capture processes entered the scintillating region.
This correlated background was estimated
using MC\@.  The normalization was constrained by the measured rate of single delayed-like candidates from this source.

\par
Table~\ref{tab:eff} is a summary of the absolute efficiencies and the systematic uncertainties. The uncertainties of the absolute efficiencies are correlated among the ADs. No relative efficiency, except $\epsilon_\mu\cdot\epsilon_m$, was corrected. All differences between the functionally identical ADs were taken as uncorrelated uncertainties.

\par
The spill-in enhancement resulted when neutrons from IBD outside the target drift into the target, and was evaluated using MC\@. The spill-out deficit ($\sim$2.2\%) was included in the absolute Gd capture ratio. The Gd capture ratio was studied using \mbox{Am-C} neutron data and MC at the detector center and the spallation neutron data and was determined using IBD MC\@. Efficiencies associated with the delayed-energy, the prompt-energy, and the capture-time cuts were evaluated with MC\@. Discussion of the uncertainties in the number of target protons, live time, and the efficiency of the flasher cut can be found in Ref.~\cite{ad12}.

\begin{table}[!htb]
\begin{center}
\begin{tabular}{lrrr}
\hline
\hline
\multicolumn{4}{c}{\bf Detector} \\\hline
  & Efficiency & Correlated & Uncorrelated \\ \hline
Target Protons  &     & 0.47\% & 0.03\%  \\
Flasher cut & 99.98\% & 0.01\% & 0.01\% \\
Delayed energy cut & 90.9\% & 0.6\% & 0.12\%    \\
Prompt energy cut  & 99.88\% & 0.10\%  & 0.01\%  \\
Multiplicity cut & & 0.02\% & $<$0.01\% \\
Capture time cut  & 98.6\% & 0.12\%  &  0.01\%  \\
Gd capture ratio  & 83.8\% & 0.8\% & $<$0.1\%  \\
Spill-in     & 105.0\% & 1.5\% & 0.02\%    \\
Livetime & 100.0\% & 0.002\% & $<$0.01\% \\

\hline
Combined & 78.8\% & 1.9\% & 0.2\% \\

\hline
\hline
 \multicolumn{4}{c}{\bf Reactor} \\\hline
 \multicolumn{2}{c|}{Correlated} & \multicolumn{2}{c}{Uncorrelated} \\\hline
Energy/fission & 0.2\%  & Power  & 0.5\% \\
IBD reaction/fission & 3\%  & Fission fraction & 0.6\% \\
&& Spent fuel & 0.3\% \\
\hline
Combined & 3\%  & Combined  & 0.8\% \\\hline

\end{tabular}
\caption{Summary of absolute efficiencies, and correlated and uncorrelated systematic uncertainties. \label{tab:eff}}
\end{center}
\end{table}

\par
Uncorrelated relative uncertainties have been addressed in detail by performing a side-by-side comparison of two ADs~\cite{ad12}. The IBD $n$Gd energy peaks for all six ADs were reconstructed to $8.05\pm 0.04$~MeV\@. The relative energy scale between ADs was established by comparing the $n$Gd peaks of the IBD- and spallation-neutrons, and alpha-particles in the \mbox{Gd-LS}\@.
Both energy-reconstruction approaches yielded a 0.5\% uncorrelated energy-scale uncertainty for all six ADs. The relative uncertainty in efficiency due to the $E_d$ cut was determined to be 0.12\% using data.
By measuring the difference in the neutron capture time of each AD, from which
the Gd-concentration can be calculated, the relative uncertainty in the fraction of neutrons captured on Gd (the Gd capture ratio) was found to be
$<$0.1\%. All other relative uncertainties were $O(0.01\%)$ and the combined uncertainty was 0.2\%. Independent analyses obtained similar results on the background and relative uncertainties.

\par
This analysis was independent of reactor flux models. The \nuebar\ yield per fission~\cite{declais} was not fixed when determining $\sin^22\theta_{13}$. Whether we used the conventional ILL fluxes~\cite{illschr,illvonf,illhahn,vogel238} (2.7\% uncertainty) or the recently calculated fluxes~\cite{huber,mueller} (3.1\% uncertainty) had little impact on the results. The thermal energy released per fission is given in Ref.~\cite{kopeikin}. Non-equilibrium corrections for long-lived isotopes were applied following Ref.~\cite{mueller}. Contributions from spent fuel~\cite{anfp,zhoub} ($\sim$0.3\%) were included as an uncertainty.

\par

\begin{figure}[htb]
\includegraphics[width=\columnwidth]{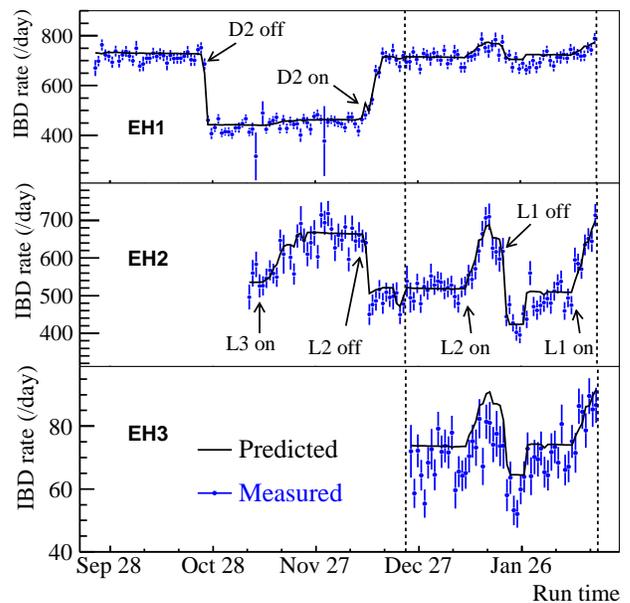}
\caption{Daily average measured IBD rates per AD in the three experimental halls as a function of time. Data between the two vertical dashed lines were used in this analysis. The black curves represent no-oscillation predictions based on reactor flux analyses and detector simulation for comparison. The predictions have been corrected with the best-fit normalization parameter in determining $\sin^22\theta_{13}$.  \label{fig:flux}}
\end{figure}

Thermal-power data provided by the power plant carry an uncertainty of 0.5\% per core~\cite{cjflux,kme,helishi} that we conservatively treat as uncorrelated. The fission fractions were also provided for each fuel cycle as a function of burn-up, with a $\sim$5\% uncertainty from validation of the simulation~\cite{sciencecode,appollo}. A DRAGON~\cite{dragon} model was constructed to study the correlation among the fission rates of isotopes. The uncertainties of the fission fraction simulation resulted in a 0.6\% uncorrelated uncertainty of the \nuebar\ yield per core. The baselines have been surveyed with GPS and modern theodolites to a precision of 28 mm. The uncertainties in the baseline  and the spatial distribution of the fission fractions in the core had a negligible effect to the results.
Fig.~\ref{fig:flux} presents the background-subtracted and efficiency-corrected IBD rates in the three EHs. Relative reactor flux predictions are shown for comparison.

\par
The \nuebar\ rate in the far hall was predicted with a weighted combination of the two near hall measurements assuming no oscillation. The weights were determined by the thermal power of each reactor and its baseline to each AD. We observed a deficit in the far hall, expressed as a ratio of observed to expected events,
\begin{equation}
R=0.940\pm 0.011({\rm stat}) \pm 0.004({\rm syst})\,.\nonumber
\end{equation}
\noindent In addition, the residual reactor-related uncertainties were found to be 5\% of the uncorrelated uncertainty of a single core.

\par
The value of $\sin^22\theta_{13}$ was determined with a $\chi^2$ constructed with pull terms accounting for the correlation of the systematic errors~\cite{stump},
\begin{eqnarray}  \label{eqn:chi2}
 \chi^2 &=&
 \sum_{d=1}^{6}
 \frac{\left[M_d-T_d\left(1+  \varepsilon
 + \sum_r\omega_r^d\alpha_r
 + \varepsilon_d\right) +\eta_d\right]^2}
 {M_d+B_d}  \nonumber \\
 &+&
 \sum_r\frac{\alpha_r^2}{\sigma_r^2}
 + \sum_{d=1}^{6} \left(
 \frac{\varepsilon_d^2}{\sigma_d^2}
 + \frac{\eta_d^2}{\sigma_{B}^2}
 \right)
 \,,
\end{eqnarray}
where $M_d$ are the measured IBD events of the $d$-th AD with backgrounds subtracted, $B_d$ is the corresponding background, $T_d$ is the prediction from neutrino flux, MC, and neutrino oscillations~\cite{fullosc},
$\omega_r^d$ is the fraction of IBD contribution of the $r$-th reactor to the $d$-th AD determined by baselines and reactor fluxes. The uncertainties are listed in Table~\ref{tab:eff}.  The uncorrelated reactor uncertainty is $\sigma_r$ (0.8\%), $\sigma_d$ (0.2\%) is the uncorrelated detection uncertainty, and $\sigma_{B}$ is the background uncertainty listed in Table~\ref{tab:ibd}. The corresponding pull parameters are ($\alpha_r, \varepsilon_d, \eta_d$). The detector- and reactor-related correlated uncertainties were not included in the analysis; the absolute normalization $\varepsilon$ was determined from the fit to the data. The best-fit value is
\begin{equation}
\sin^22\theta_{13}=0.092\pm 0.016({\rm stat})\pm0.005({\rm syst})
\nonumber
\end{equation}
with a $\chi^2$/NDF of 4.26/4. All best estimates of pull parameters are within its one standard deviation based on the corresponding systematic uncertainties. The no-oscillation hypothesis is excluded at 5.2 standard deviations.

\par
The accidental backgrounds were uncorrelated while the \mbox{Am-C} and (alpha,n) backgrounds were correlated among ADs.  The fast-neutron and $^9$Li/$^8$He backgrounds were site-wide correlated. In the worst case where they were correlated in the same hall and uncorrelated among different halls, we found the best-fit value unchanged while the systematic uncertainty increased by 0.001.

\par
\begin{figure}[htb]
\includegraphics[width=\columnwidth]{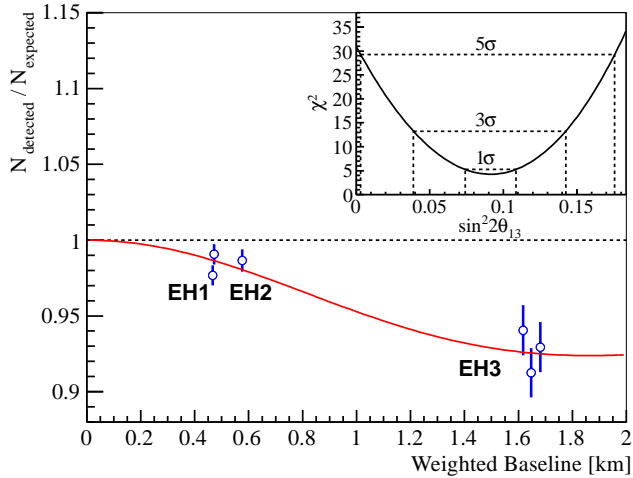}
\caption{
Ratio of measured versus expected signal in each detector, assuming no oscillation. The error bar is the uncorrelated uncertainty of each AD, including statistical, detector-related, and background-related uncertainties. The expected signal is corrected with the best-fit normalization parameter. Reactor and survey data were used to compute the flux-weighted average baselines. The oscillation survival probability at the best-fit value is given by the smooth curve. The AD4 and AD6 data points are displaced by -30 and +30 m for visual clarity. The $\chi^2$ versus $\sin^22\theta_{13}$ is shown in the inset. \label{fig:osc}}
\end{figure}

Fig.~\ref{fig:osc} shows the measured numbers of events in each detector, relative to those expected assuming no oscillation. The 6.0\% rate deficit is obvious for EH3 in comparison with the other EHs, providing clear evidence of a non-zero $\theta_{13}$.  The oscillation survival probability at the
best-fit values is given by the smooth curve.  The $\chi^2$
versus sin$^22\theta_{13}$ is shown in the inset.

\par
\begin{figure}[htb]
\includegraphics[width=\columnwidth]{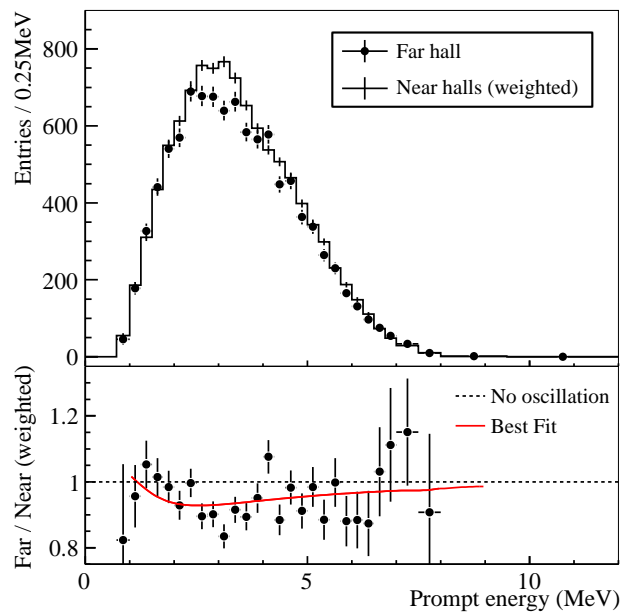}
\caption{Top: Measured prompt energy spectrum of the far hall (sum of three ADs) compared with the no-oscillation prediction from the measurements of the two near halls. Spectra were background subtracted. Uncertainties are statistical only. Bottom: The ratio of measured and predicted no-oscillation spectra. The red curve is the best-fit solution with $\sin^22\theta_{13}=0.092$ obtained from the rate-only analysis. The dashed line is the no-oscillation prediction.  \label{fig:spec}}
\end{figure}

The observed \nuebar\  spectrum in the far hall is compared to a prediction based on the near hall measurements in Fig.~\ref{fig:spec}. The disagreement of the spectra provides further evidence of neutrino oscillation. The ratio of the spectra is consistent with the best-fit oscillation solution of $\sin^22\theta_{13}=0.092$ obtained from the rate-only analysis~\cite{fnshape}.

\par
In summary, with a 43,000 ton-GW$_{\rm th}$-day livetime exposure, 10,416 reactor antineutrinos were observed at the far hall. Comparing with the prediction based on the near-hall measurements, a deficit of 6.0\% was found.
A rate-only analysis yielded $\sin^22\theta_{13}=0.092\pm 0.016({\rm stat})\pm0.005({\rm syst})$.
The neutrino mixing angle $\theta_{13}$ is non-zero with a significance of 5.2 standard deviations.

\par
The Daya Bay experiment is supported in part by the Ministry of Science and Technology of China, the United States Department of Energy,  the Chinese Academy of Sciences, the National Natural Science Foundation of China, the Guangdong provincial government, the Shenzhen municipal government,  the China Guangdong Nuclear Power Group, Shanghai Laboratory for Particle Physics and Cosmology, the Research Grants Council of the Hong Kong Special Administrative Region of China, University Development Fund of The University of Hong Kong, the MOE program for Research of Excellence at National Taiwan University, National Chiao-Tung University, and NSC fund support from Taiwan, the U.S. National Science Foundation, the Alfred~P.~Sloan Foundation, the Ministry of Education, Youth and Sports of the Czech Republic, the Czech Science Foundation, and the Joint Institute of Nuclear Research in Dubna, Russia. We thank Yellow River Engineering Consulting Co., Ltd.\ and China railway 15th Bureau Group Co., Ltd.\ for building the underground laboratory. We are grateful for the ongoing cooperation from the China Guangdong Nuclear Power Group and China Light~\&~Power Company.


\begin{thebibliography}{99}
\bibitem{fullosc} The survival probability used in the $\chi^2$ was\\
$P_{sur} = 1 - \sin^22\theta_{13}\sin^2(1.267\Delta m^2_{31}L/E) - \\
\cos^4\theta_{13}\sin^22\theta_{12}\sin^2(1.267\Delta m^2_{21}L/E)$
where, $\\ \Delta m^2_{31} = 2.32\times10^{-3} {\rm eV}^2,
\sin^22\theta_{12} = 0.861^{+0.026}_{-0.022}$, and
$\\ \Delta m^2_{21} = 7.59^{+0.20}_{-0.21}\times 10^{-5} {\rm eV}^2$. The uncertainty in $\Delta m_{31}^2$~\cite{minosdm} has not been included in the fit. The fit $\sin^22\theta_{13}$ will change by +0.0007 and -0.0004 when $\Delta m^2_{31}$ changes by one standard deviation.
\bibitem{fnshape} Without correcting for the non-linearity of the detector response, we have performed a preliminary shape analysis that yielded a consistent result for $\sin^22\theta_{13}$.
\end{thebibliography}
\end{document}